\documentclass[12pt]{article}

\usepackage{framed}
\usepackage{cancel}
\usepackage{braket}
\usepackage{graphicx}
\usepackage{mathrsfs}
\usepackage{amsmath,amssymb}
\usepackage{listings}
\usepackage[most]{tcolorbox}
\usepackage{inconsolata}
\usepackage{times}
\usepackage{hyperref}
\usepackage[english]{babel}
\usepackage[utf8]{inputenc}
\usepackage[normalem]{ulem}

\definecolor{blue-violet}{rgb}{0.54, 0.17, 0.89}
\definecolor{PineGreen}{cmyk}{0.92, 0, 0.59, 0.25}
\definecolor{Gray}{cmyk}{0, 0, 0, 0.50}

\newtcblisting[auto counter]{codeprint}[2][]{sharp corners,
    fonttitle=\bfseries, colframe=gray, listing only,
    listing options={basicstyle=\ttfamily,language=C++},
    title=C\'odigo \thetcbcounter: #2, #1}

\numberwithin{equation}{section}

\newcounter{lastnote}

\title{First-order Lagrangian and Hamiltonian of Lovelock gravity}
\author{Pablo Guilleminot$^\S$, F\'elix-Louis Juli\'e$^\P$, Nelson Merino$^\ddag$,\\
\& Rodrigo Olea$^\S$  \smallskip \\
$^\S${\small \emph{Departamento de Ciencias F\'isicas, Universidad
Andres Bello,}}\\
{\small \emph{Sazi\'e 2212, Piso 7, Santiago, Chile. \smallskip }}\\
$^\P${\small \emph{Department of Physics and Astronomy, Johns %
Hopkins University,}}\\
{\small \emph{ \smallskip 3400 N. Charles Street, Baltimore, MD 21218, USA.}}\\
$^\ddag${\small \emph{Instituto de Ciencias Exactas y Naturales (ICEN), Facultad de Ciencias,}}\\
{\small \emph{ \smallskip Universidad Arturo Prat, 1110939 Iquique, Chile.}}\\}

\date{\today}
\begin{document}

\baselineskip20pt

\maketitle

\begin{abstract} 
\medskip
Based on the insight gained by many authors over the years on the structure of the Einstein-Hilbert, Gauss-Bonnet and Lovelock gravity Lagrangians, we show how to derive --in an elementary fashion-- their first-order, generalized ``ADM" Lagrangian and associated Hamiltonian. To do so, we start from the
Lovelock Lagrangian supplemented with the Myers boundary term, which guarantees a Dirichlet variational principle with a surface term of the form $\pi^{ij}\delta h_{ij}$, where $\pi^{ij}$ is the canonical momentum 
conjugate to the boundary metric $h_{ij}$. Then, the first-order Lagrangian density is obtained either by integration of $\pi^{ij}$ over the metric derivative $\partial_wh_{ij}$ normal to the boundary, or by rewriting the Myers term as a bulk term. 
\end{abstract}

\section*{Introduction}

The General Relativity (GR), Gauss-Bonnet (GB) and more generally Lovelock \cite{LanLov} Lagrangians, being (quasi) linear in the second derivatives of the metric, yield second-order field equations (see e.g. \cite{DM} for a review).

There must hence exist \textit{first-order} Lagrangians,  which do not depend on the metric's second derivative normal to a foliation, and which differ from Lovelock's by adding adequate boundary terms, so that they produce the same dynamics but with Dirichlet boundary conditions. 

In general relativity, a boundary term to be added to the Einstein-Hilbert Lagrangian to yield a Dirichlet variational principle was proposed by Gibbons, Hawking \cite{GHY} and York \cite{York:1972sj} (GHY). Its generalization to GB
and Lovelock theories was obtained by Myers \cite{Myers}, see also \cite{MuHo,Davis,GravWill,MisOl}.

In general relativity, a well-known first-order Lagrangian is that of Arnowitt, Deser, and Misner (ADM), which is written (as well as the corresponding Hamiltonian) in a 1+3 form in terms of the extrinsic and intrinsic curvatures of a spacetime foliation \cite{ADM,ADM2}.
The GB and Lovelock first-order Lagrangians (and corresponding Hamiltonian) generalizing ADM's were found by Teitelboim and Zanelli \cite{TeiZan,TZ}.

In this paper, we will obtain the Teitelboim-Zanelli Lagrangian and Hamiltonian in two different straightforward manners.
We shall first illustrate the methods on the (nowadays) simple case of general relativity, and then generalize the procedure to all Lovelock Lagrangians.

\section{The crux of the method}

\subsection{The example of point mechanics\label{PMsubsection}}

Consider a particle with position $q(t)$ described by the action
\begin{equation}  \label{eq1.1}
I=\int_{t_i}^{t_f} dt\,\! L\quad\text{with}\quad L(q,\dot q,\ddot q)=\ell(q,\dot q)+\ddot q f(q,\dot q)\,,
\end{equation}
where a dot denotes a derivative with respect to time $t$.
The variation of $I$ upon an infinitesimal variation $\delta q(t)$ of the path $q(t)$ reads
\begin{equation} \label{eq2}
\delta I=\int_{t_i}^{t_f}\! dt\,\delta q\left[B(q,\dot q)-\ddot q A(q,\dot q)\right]+\left[\delta q\left(\frac{\partial \ell}{\partial\dot q}-\dot q\frac{\partial f}{\partial q}\right)+\delta\dot q\, f\right]_{t_i}^{t_f}\,.
\end{equation}

The issue with $I$ is that its variation $\delta I$ cannot be made to vanish for an arbitrary $\delta q(t)$ between $t_i$ and $t_f$. Indeed, the vanishing of the boundary terms necessitates fixing 4 constants (to wit the positions  and velocities of the particle at $t_i$ and $t_f$ so that $\delta q|_{t_i}=\delta q|_{t_f}=\delta\dot q|_{t_i}=\delta\dot q_{t_f}=0$). These conditions are incompatible with the fact that the solutions of the equation of motion ($B-\ddot q\, A=0$), which is second order since $L$ is (quasi) linear in the acceleration $\ddot q$, depend on 2 integration constants only.\footnote{For completeness~: $A(q,\dot q)=\frac{\partial^2\ell}{\partial\dot q^2}-\dot q\frac{\partial^2f}{\partial q\partial\dot q}-2\frac{\partial f}{\partial q}$ and $B(q,\dot q)=\frac{\partial}{\partial q}\left(\ell-\dot q\frac{\partial \ell}{\partial\dot q}+\dot q^2\frac{\partial f}{\partial q}\right)$.}

Now, it must be possible to build an ordinary, first-order Lagrangian $L_{1}(q,\dot q)$ and associated action $I_{1}$ which yield a second order equation of motion when imposing $\delta I_{1}=0$ for Dirichlet boundary conditions (that is, by fixing $\delta q|_{t_i}=\delta q|_{t_f}=0$ only). In order to give the same equation of motion as $L$,  $L_{1}(q,\dot q)$ is taken to differ from $L$ by the substraction of a total time derivative of some function $F(q,\dot q)$~:
\begin{equation} \label{eq3}
L_{1}(q,\dot q)=L-\frac{dF(q,\dot q)}{dt}\quad,\quad I_{1}=\int_{t_i}^{t_f}\! dt\, L_{1}=I-\left[F(q,\dot q)\right]_{t_i}^{t_f}\,.
\end{equation}

A simple route to obtain $L_{1}$ is to compute the surface terms in the variation of the action. We have, on-shell, that is when the equation of motion is satisfied,
\begin{eqnarray} \label{bulkCM}
\delta I_{1}=\left[\delta q\left({\partial \ell\over\partial\dot q}-\dot q{\partial f\over\partial q}-\frac{\partial F}{\partial q}\right)+\delta\dot q\left( f-\frac{\partial F}{\partial\dot q}\right)\right]_{t_i}^{t_f}\,,
\end{eqnarray}
where we have used (\ref{eq2}). 
The vanishing of the coefficient of $\delta\dot q$ in (\ref{bulkCM}) gives the function $F$,
\begin{equation} \label{eq5}
F=\int\!d\dot q\,f(q,\dot q)\,.
\end{equation}
If we then identify the coefficient of $\delta q$ to the canonical momentum (see e.g. \cite{LLbook})
\begin{equation}
p=\frac{\partial L_{1}}{\partial\dot q} \,,
\end{equation}
 $L_1$ is obtained by a simple integration with respect to the velocity $\dot q$:
 \begin{equation}\label{eq5.5}
L_1=\ell(q,\dot q)-\dot q{\partial F\over\partial q}
 \end{equation}
 with $F$ given by Eq. (\ref{eq5}).
 
Another way, even simpler in this case, to obtain $L_{1}$ is to lift $F$ to the bulk (a procedure which we shall refer to as \textit{bulkanization} below), and write, using (\ref{eq3}) and (\ref{eq1.1}):
\begin{eqnarray} 
I_{1}&\equiv& \int_{t_i}^{t_f}\! dt\, L_{1}(q,\dot q)\notag \\
&=&\int_{t_i}^{t_f}\! dt \left[L-\frac{dF}{dt}\right] \notag\\
&=&\int_{t_i}^{t_f}\! dt\left[\ell(q,\dot q)-\dot q{\partial F\over\partial q}+\ddot q\left(f-{\partial F\over\partial\dot q}\right)\right]\,,
\end{eqnarray}
which yields back (\ref{eq5.5}), using (\ref{eq5}):\footnote{It is an exercise to check that the equation of motion derived from $L_{1}$  is the same as that derived from $L$~: $\dot p-{\partial L_{1}\over\partial q} = \ddot q\,A -B$, with $A$ and $B$ given in footnote 1. As for the Hamiltonian $H= p \dot q - L_1$, it cannot, in general, be written explicitely in terms of $q$ and $p$ unless $p= p(q, \dot q)$ can be  inverted explicitely to give $\dot q= \dot q(q,p)$. Hence it cannot be shown explicitely that the Hamilton equations yield back the Euler-Lagrange equations derived from $L_1$.}

\subsection{Two routes to the first-order Lagrangian of GR}  \label{Dir_in_GR}

Let us first recall how the Gibbons-Hawking-York (GHY) boundary term is obtained. Consider, in some coordinate system $x^\mu$ labelling the points of a $D$-dimensional pseudo-Riemannian manifold $\mathcal M$ (Greek indices run from $0$ to $D-1$; see Appendix \ref{GC} for conventions), the GR action
\begin{equation}
I_{\rm GR}=\int\limits_{\mathcal M}d^Dx\sqrt{-g}R\ .\label{bulkActionGR}
\end{equation}
This action depends linearly on the second derivatives of the field variables $g_{\mu\nu}$, and its variation reads:
\begin{equation}
\delta I_{\rm GR}=\int\limits_{\mathcal M}d^Dx\sqrt{-g}\left(G_{\mu\nu}\delta g^{\mu\nu}+\nabla_\mu V_{\rm GR}^\mu\right)\ ,\label{covariantVariationGR}
\end{equation}
where $G_{\mu\nu}$ is the Einstein tensor. The second term on the r.h.s. of (\ref{covariantVariationGR}) is the covariant divergence of  the four-vector
\begin{align} \label{bdrygr}
V_{\rm GR}^\mu&=g^{\alpha\beta}\delta\Gamma^\mu_{\alpha\beta}-g^{\mu\alpha}\delta\Gamma^{\beta}_{\alpha\beta} \ ,
\end{align}
which can be evaluated, using Gauss' theorem, on the $d=D-1$ dimensional boundary $\partial\mathcal M$ of $\mathcal M$.

Let us choose for simplicity a Gaussian coordinate system $x^\mu=\{w,x^i\}$ (Latin indices run from 1 to $d=D-1$), such that $w$ is constant on $\partial\mathcal M$:
\begin{equation} \label{mmetric}
ds^2=\epsilon\,N(w)^2 dw^2+h_{ij}(w,x^k)dx^i dx^j\ ,
\end{equation} 
with $\epsilon=-1$ if $\partial\mathcal M$ is spacelike and  $\epsilon=+1$ if it is timelike, where $N(w)$ is a function of $w$ only and $h_{ij}$ are the $d(d+1)/2$ components of the induced metric on $\partial\mathcal M$, with extrinsic curvature
\begin{equation} 
K_{ij}=\frac{1}{2N}\partial_wh_{ij}\ .
\end{equation}
From now on latin indices are lowered and raised with $h_{ij}$ and its inverse $h^{ij}$.
For the gauge-fixed metric (\ref{mmetric}) we have
\begin{equation}
V^w_{\rm GR}=-\frac{\epsilon}{N}\big(K^{ij}\delta h_{ij}+2\delta K\big)\,,
\end{equation}
where $K=h^{ij}K_{ij}$, making manifest that the surface term in (\ref{covariantVariationGR}) contains variations of the normal derivative of $h_{ij}$ through $\delta K$ (the latter originates from the components (\ref{chrGN}) of $\delta \Gamma$).
 
Hence a Dirichlet action principle can be achieved if the GR action is supplemented with the GHY boundary term \cite{GHY, York:1972sj}
\begin{equation}\label{ID}
I_{\rm Dir}[g]=\int\limits_{\mathcal{M}}d^Dx\sqrt{-g}\,R +2\epsilon\!\int\limits_{\partial\mathcal{M}}d^dx\sqrt{|h|}K\,,
\end{equation}
since the variation of this action gives, on-shell (that is, when $G_{\mu\nu}=0$ in vacuum),
\begin{equation} \label{adaptedVariationGR}
\delta I_{\rm Dir}=\int\limits_{\partial\mathcal{M}}d^d x%
\,\pi^{ij}\delta h_{ij}\,,
\end{equation}
where 
\begin{equation} \label{pi1}
\pi^{ij}=\epsilon \sqrt{|h|}(Kh^{ij}-K^{ij})\ ,
\end{equation}
and vanishes imposing Dirichlet boundary conditions:
$\delta h_{ij}\vert_{\partial\mathcal M}=0$.

The action principle above can be associated to a first-order bulk functional,
\begin{align}
I_{1}=\int\limits_{\mathcal M}d^Dx\, \mathcal L_{1}\, .\label{actionADM}
\end{align}
Indeed, in a Gaussian frame (\ref{mmetric}) which foliates $\mathcal M$ with constant-$w$ surfaces $\Sigma_w$, $\mathcal L_{1}$ can be obtained by identifying 
$\pi^{ij}$ given by Eq.(\ref{pi1}) as the canonical momentum density conjugate to $h_{ij}$, i.e.,
\begin{align}
\frac{\partial \mathcal L_{1}}{\partial(\partial_w h_{ij})}&=\pi^{ij}\,.\label{eqDiffLagrangianGR}
\end{align}
Integrating $\pi^{ij}$ with respect to $\partial_w h_{ij}=2NK_{ij}$ gives
\begin{align}
\mathcal L_{1}&=N\sqrt{|h|}\bigg(\epsilon \left(K^2-K^{ij}K_{ij}\right)+r(h_{ij}, \partial_kh_{ij}, \partial_{k}\partial_{l}h_{ij}) \bigg)\, ,
\end{align}
where the integration constant $r(h_{ij}, \partial_kh_{ij}, \partial_{k}\partial_{l}h_{ij})$ must identify to the part of the Hilbert Lagrangian which only depends on the intrinsic geometry of the surfaces $\Sigma_w$, i.e. $\bar R$, where a bar stands for quantities built out of $h_{ij}$ only\footnote{By \textit{intrinsic geometry}, we refer to quantities built out of $h_{ij}$ and its tangential derivatives $\partial_{k}h_{ij}$ and $\partial_{k}\partial_{l}h_{ij}$ only.}
\begin{align}
\mathcal L_{1}&=N\sqrt{|h|}\bigg(\bar R+\epsilon \left(K^2-K^{ij}K_{ij}\right)\bigg)\nonumber\\
&=\mathcal L_{\rm ADM}\,.\label{lagrangianADM_GR}
\end{align}
This is the celebrated ADM Lagrangian density \cite{ADM,ADM2} written here in Gaussian coordinates.

Let us show now that the same first-order (in the normal derivative) Lagrangian density can be obtained by the bulkanization of the GHY term. Define the closed boundary by the union $\partial\mathcal M=\Sigma_{w_i}\cup\Sigma_{w_f}\cup\mathcal C$ of the surfaces $w=w_i$ and $w=w_f$ and their complement $\mathcal C$, and rewrite the GHY contributions from $\Sigma_{w_i}$ and $\Sigma_{w_f}$ in (\ref{ID}) as the integral of $2\epsilon\,\partial_w\big(\sqrt{|h|}K\big)$ over the bulk. Using the Gauss-Codazzi-Mainardi relation (\ref{Rscalar}), we then have
\begin{equation} \label{a1}
\sqrt{-g}R+2\epsilon\,\partial_w\big(\sqrt{|h|}K\big)=\sqrt{-g}\big[%
\bar R-\epsilon \big(K^2+K^i_jK^j_i%
\big)\big]+2\epsilon\,\partial_w\big(\sqrt{|h|}\big)K\,.
\end{equation}
Since moreover
$\partial_w\sqrt{|h|}=NK\sqrt{|h|}$,
we obtain 
\begin{eqnarray} \label{a2}
\sqrt{|h|}R+2\epsilon\,\partial_w\big(\sqrt{|h|}K\big)&=&N\sqrt{|h|}\bigg(\bar R+\epsilon \left(K^2-K^{ij}K_{ij}\right)\bigg)\notag \\
&=&\mathcal{L}_{\rm ADM}\,.
\end{eqnarray}
The bulkanized GHY terms on $\Sigma_{w_i}$ and $\Sigma_{w_f}$ cancel out with the second normal derivative in Eq. (\ref{a1}) that comes from $R^{wi}_{wj}$, see (\ref{Rwiwj}), so that the  resulting Lagrangian is of first order. As for the GHY defined on the complement $\mathcal C$, it can be discarded for our purposes (but is essential to define the ADM mass \cite{poisson}).

Finally, the dependence on the $D=d+1$ extra components of the spacetime metric $g_{\mu\nu}$ can be reinstated using the ADM metric decomposition
\begin{align}
ds^2=\epsilon N^2dw^2+h_{ij}(dx^i+N^idw)(dx^j+N^jdw)\,,\label{ADMcoordinates}
\end{align}
where $N(w,x^i)$ is the lapse and $N^i(w,x^j)$ is the shift. The extrinsic curvature is then redefined as
\begin{align}\label{Kadm}
K_{ij}&=\frac{1}{2N}(\partial_w h_{ij}-\bar\nabla_iN_j-\bar\nabla_jN_i)\,,
\end{align}
with $\bar\nabla_i$ the covariant derivative associated to $h_{ij}$.

It can be explicitely checked that variations with respect to $N$, $N^i$ and $h_{ij}$ of $\mathcal{L}_{\rm ADM}$ yield respectively the constraints $G^w_w=0$, $G^i_w=0$ and the dynamical component $G^i_j=0$ of the equations of motion written in Gaussian coordinates.

\section{The first-order Lagrangian of Lovelock gravity}
\subsection{Dirichlet principle for Lovelock gravity}

As shown by Myers \cite{Myers}, the Dirichlet action for a generic Lovelock theory is given by
\begin{equation}
I_{\mathrm{Dir}}=\sum_{p=0}^{[\frac{D-1}{2}]}\alpha
_{p}\Bigg(\int\limits_{\mathcal{M}}d^{D}x\mathcal{L}^{(p)}-\int\limits_{\partial
\mathcal{M}}d^{d}x\beta^{(p)}\Bigg)\,,\label{IDL}%
\end{equation}
where $[(D-1)/2]$ is the integer part of $(D-1)/2$, where\footnote{In even dimensions, the term $p=D/2$ is topological, and it does not contribute to the field equations.}
\begin{equation}
\mathcal{L}^{(p)}=\frac{1}{2^{p}}\sqrt{-g}\delta_{\lbrack\nu_{1}\cdots\nu_{2p}%
]}^{[\mu_{1}\cdots\mu_{2p}]}R_{\mu_{1}\mu_{2}}^{\nu_{1}\nu_{2}}\cdots R_{\mu
_{2p-1}\mu_{2p}}^{\nu_{2p-1}\nu_{2p}}\,,\label{Lp}%
\end{equation}
is of degree $p$ in the curvature, and where
\begin{equation} 
\delta _{\left[ \mu _{1}\cdots \mu _{2p}\right] }^{\left[ \nu _{1}\cdots \nu
_{2p}\right] }\equiv\left\vert
\begin{array}{cccc}
\delta _{\mu _{1}}^{\nu _{1}} & \delta _{\mu _{1}}^{\nu _{2}} & \cdots &
\delta _{\mu _{1}}^{\nu _{2p}} \\
\delta _{\mu _{2}}^{\nu _{1}} & \delta _{\mu _{2}}^{\nu _{2}} &  & \delta
_{\mu _{2}}^{\nu _{2p}} \\
\vdots &  & \ddots &  \\
\delta _{\mu _{2p}}^{\nu _{1}} & \delta _{\mu _{2p}}^{\nu _{2}} &  &
\delta _{\mu _{2p}}^{\nu _{2p}}%
\end{array}%
\right\vert\,,
\end{equation} 
is the generalized Kronecker delta of rank $2p$, which is antisymmetric under exchange of its upper (and lower) indices. In our conventions (see Appendix \ref{GC}), the dimension of $\alpha_p$ is $[{\rm length]^{2p-2}}$. The corresponding Myers boundary terms are given by \cite{Myers,Davis}
\begin{align}
\beta^{(p)} &  =-2\epsilon p\sqrt{|h|}\int_{0}^{1}\!ds\,\delta_{\lbrack j_{1}\cdots
j_{2p-1}]}^{[i_{1}\cdots i_{2p-1}]}K_{i_{1}}^{j_{1}}\left(  \frac{1}{2}\bar
{R}_{i_{2}i_{3}}^{j_{2}j_{3}}-s^{2}\epsilon K_{i_{2}}^{j_{2}}K_{i_{3}}^{j_{3}%
}\right)  \times\cdots\nonumber\label{B2p}\\
& \hspace{150pt} \cdots\times\left(  \frac{1}{2}\bar{R}_{i_{2p-2}i_{2p-1}}^{j_{2p-2}%
j_{2p-1}}-s^{2}\epsilon K_{i_{2p-2}}^{j_{2p-2}}K_{i_{2p-1}}^{j_{2p-1}}\right)
\,.
\end{align}
For its rewriting as the covariant derivative of a $D-$vector, see also \cite{DMO} or \cite{Coll} which involve, respectively, the introduction of a background metric or an extra vector field which identifies to the normal $n$ on $\partial\mathcal M$.
In our conventions we have $\alpha_{0}=-2\Lambda$ and
$\alpha_{1}=1$.

The variation of Eq. (\ref{IDL}) reads 
\begin{equation} \label{ILV}
\delta I_{\mathrm{Dir}}=\int\limits_{\mathcal{M}}%
d^{D}x\sqrt{-g}\mathcal{E}^{\mu\nu}\delta g_{\mu\nu}+%
\int\limits_{\partial\mathcal{M}}d^{d}x\,\pi^{ij}\delta h_{ij}\,,
\end{equation}
with 
\begin{equation} \label{canmom}
\pi^{ij}=\sum_{p=0}^{[\frac{D-1}{2}]}\alpha_{p}\pi^{ij}_{(p)}\,,
\end{equation}
where, from each $p$th Lovelock density, one obtains
\begin{align}
\pi^{ij}_{(p)} &  =p\epsilon\sqrt
{|h|}\int_{0}^{1}\!ds\,\delta_{[kj_{1}\cdots j_{2p-1}]}^{[ii_{1}\cdots
i_{2p-1}]}h^{kj}K_{i_{1}}^{j_{1}}\Bigg(\frac{1}{2}\bar{R}_{i_{2}i_{3}}^{j_{2}%
j_{3}}-s^{2}\epsilon K_{i_{2}}^{j_{2}}K_{i_{3}}^{j_{3}}\Bigg)  \times
\cdots\nonumber\\
& \hspace{150pt} \cdots\times\Bigg(  \frac{1}{2}\bar{R}_{i_{2p-2}i_{2p-1}}^{j_{2p-2}%
j_{2p-1}}-s^{2}\epsilon K_{i_{2p-2}}^{j_{2p-2}}K_{i_{2p-1}}^{j_{2p-1}}
\Bigg)\,.\label{Pi_new}%
\end{align}
As for the Lovelock tensor $\mathcal{E}_{\nu}^{\mu}$, it reads
\begin{equation}
\mathcal{E}_{\nu}^{\mu}=\sum_{p=0}^{[\frac{D-1}{2}]}\alpha_{p}\,\mathcal{E}_{(p)\nu}^{\mu}\label{eom}%
\end{equation}
with
\begin{equation}
\mathcal{E}_{(p)\nu}^{\mu}=-\frac{1}{2^{p+1}}%
\delta_{\lbrack\nu\nu_{1}\cdots\nu_{2p}]}^{[\mu\mu_{1}\cdots\mu_{2p}]}%
R_{\mu_{1}\mu_{2}}^{\nu_{1}\nu_{2}}\cdots R_{\mu_{2p-1}\mu_{2p}}^{\nu
_{2p-1}\nu_{2p}}\,.\label{eomP}%
\end{equation}
Note that in the boundary term of (\ref{ILV}) we omitted the divergence of a $d$-vector $\bar\nabla_i W^i$ since its integration on the closed boundary $\partial\mathcal M$ vanishes (see, e.g., \cite{DM}; see also \cite{Chak} for its explicit expression).

The addition of a topological term in even dimensions cannot induce an associated canonical momentum $\pi^{ij}_{(D/2)}$. This can be seen from the anti-symmetric structure of the indices in the canonical momentum in Eq.(\ref{Pi_new}). In the critical space-time dimension, the canonical momentum is constructed with a Kronecker delta of rank $D$ at the boundary, a fact that makes it identically zero.\footnote{In gravity theories with AdS asymptotics, topological terms do play an essential role in the renormalization of the action and its variation (see, e.g., \cite{AMOP}). The corresponding coupling, however, is not arbitrary, but fixed by the boundary dynamics.}

The action (\ref{IDL}) yields a Dirichlet variational principle. In other words, the Myers boundary terms are the analogues of the function $F$, given by (\ref{eq5}), in the mechanical problem we treated in section \ref{PMsubsection}.

\subsection{Two routes to the first-order Lagrangian for Lovelock gravity}

\textbf{Integration of $\bf\pi^{ij}$.} As explicitly worked out above on the example of GR, we can now
construct the first-order Lagrangian density by identifying the tensor density
(\ref{Pi_new}) as the associated canonical momentum:
\begin{equation} \label{intpi}
\frac{\partial\mathcal{L}^{(p)}_{\mathrm{ADM}}}{\partial(\partial_{w}h_{ij})}%
=\pi^{ij}_{(p)}\,.
\end{equation}

Substituting $\partial_{w}h_{ij}=2NK_{ij}$ above and integrating the canonical momentum as a polynomial of the extrinsic curvature yields the generalization of the ADM Lagrangian density to Lovelock theories, after proper inclusion of the lapse and shift:

\begin{align} \label{LpINT}
\mathcal{L}_{\mathrm{ADM}}^{(p)}  & =Nr(h_{ij}, \partial_kh_{ij}, \partial_{k}\partial_{l}h_{ij})+2p\epsilon N\sqrt{|h|}\int_{0}%
^{1}ds(1-s)\delta_{\lbrack j_{1}\cdots j_{2p}]}^{[i_{1}\cdots i_{2p}]}%
K_{i_{1}}^{j_{1}}\times \notag \\
& \times K_{i_{2}}^{j_{2}}\left(  \frac{1}{2}\bar{R}_{i_{3}i_{4}}^{j_{3}j_{4}%
}-s^{2}\epsilon K_{i_{3}}^{j_{3}}K_{i_{4}}^{j_{4}}\right)  \times\cdots
\times\left(  \frac{1}{2}\bar{R}_{i_{2p-1}i_{2p}}^{j_{2p-1}j_{2p}}%
-s^{2}\epsilon K_{i_{2p-1}}^{j_{2p-1}}K_{i_{2p}}^{j_{2p}}\right)  \,.
\end{align}
where $r(h_{ij}, \partial_kh_{ij}, \partial_{k}\partial_{l}h_{ij})$ is a function that does not depend on normal derivatives of the induced metric.
In view of the Gauss-Codazzi relations, the only intrinsic quantity coming from a $(d+1)$ decomposition of the Riemann tensor is $\bar{R}^{ij}_{kl}$. In other words, $r$ can only be the $p$th Lovelock density (\ref{Lp}) but computed using the induced metric, i.e. $r=\bar{\mathcal{L}}^{(p)}$ with
\begin{equation}
\bar{\mathcal{L}}^{(p)}=\frac{1}{2^{p}}\sqrt{|h|}\delta_{\lbrack j_{1}\cdots
j_{2p}]}^{[i_{1}\cdots i_{2p}]}\bar{R}_{i_{1}i_{2}}^{j_{1}j_{2}}\cdots\bar
{R}_{i_{2p-1}i_{2p}}^{j_{2p-1}j_{2p}}\,.\label{LpB}%
\end{equation}

\textbf{Bulkanization of the Myers term.} When the bulk Lagrangian density $\mathcal{L}^{(p)}$ is re-expressed in the coordinate frame (\ref{mmetric}), a term linear in the acceleration (that is, the normal derivatives of the extrinsic curvature) arises from $R^{wi}_{wj}$.
On the other hand, lifting $\beta^{(p)}$ to the bulk produces two types of contributions: i) normal derivatives of the extrinsic curvature, that eliminate the acceleration-dependent part coming from $\mathcal{L}^{(p)}$, ii) first-order normal derivatives of the induced metric, i.e. powers of the velocity. The latter contain, in particular, a term with an antisymmetric Kronecker delta with an additional pair of indices.

This task is explicitly carried out in Appendix \ref{ED}.
In doing so, it is useful to employ Eq. (\ref{ABF}) to derive the equivalent form of the Dirichlet action (in Gaussian coordinates)
\begin{align} \label{MET}
\int\limits_{\mathcal{M}}d^{D}x\Big(\mathcal{L}^{(p)}-\frac{d}{dw}\big(\beta^{(p)}\big)\Big) &  =-\int\limits_{\mathcal{M}}d^{D}x\,\mathcal{Q}^{(p)}%
+2p\epsilon N\int\limits_{\mathcal{M}}d^{D}x\sqrt{|h|}\int_{0}%
^{1}ds\,\delta_{\lbrack j_{1}\cdots j_{2p}]}^{[i_{1}\cdots i_{2p}]}K_{i_{1}%
}^{j_{1}}K_{i_{2}}^{j_{2}}\times\nonumber\\
&  \times\left(  \frac{1}{2}\bar{R}_{i_{3}i_{4}}^{j_{3}j_{4}}-s^{2}\epsilon
K_{i_{3}}^{j_{3}}K_{i_{4}}^{j_{4}}\right)  \times\cdots\times\left(  \frac
{1}{2}\bar{R}_{i_{2p-1}i_{2p}}^{j_{2p-1}j_{2p}}-s^{2}\epsilon K_{i_{2p-1}%
}^{j_{2p-1}}K_{i_{2p}}^{j_{2p}}\right)  \,\hspace{10pt}%
\end{align}
where 
\begin{equation}
\mathcal{Q}^{(p)}=-\frac{1}{2^{p}}N\sqrt{|h|}\delta_{\lbrack j_{1}\cdots j_{2p}%
]}^{[i_{1}\cdots i_{2p}]}R_{i_{1}i_{2}}^{j_{1}j_{2}}\times\cdots\times
R_{i_{2p-1}i_{2p}}^{j_{2p-1}j_{2p}}\,\label{Hp}%
\end{equation}
is $-\mathcal{L}^{(p)}$ saturated with intrinsic indices, where $R_{ijkl}$ is understood as a function of $\bar R_{ijkl}$ and $K_{ij}$, see (\ref{Rijkl}) (for a different decomposition see \cite{Paddy}).  We note that $\mathcal{Q}^{(p)}$ is also proportional to the $w$-$w$ component of the $p$th Lovelock tensor $\mathcal E^{\mu}_{(p)\nu}$, see (\ref{eomP}).

Using Eq. (\ref{MET}) and the Gauss-Codazzi relations to express
$\mathcal{Q}^{(p)}$ in terms of the intrinsic curvature with the identity
\begin{equation}
(x+y)^{p}=x^{p}+2py\int_{0}^{1}ds\,s(x+s^{2}y)^{p-1}\,,
\end{equation}
we can rewrite the Dirichlet action (\ref{IDL}) purely as a functional of $h_{ij}$,
$K_{ij}$ and $R_{ijkl}$ (or $\bar{R}_{ijkl}$) to obtain
\begin{align}
I_{\rm ADM}[h,K,\bar{R}]=\int\limits_{\mathcal{M}}d^{D}x\,\mathcal{L}_{\rm ADM}=\int\limits_{\mathcal{M}}d^{D}x\sum_{p=0}^{[\frac{D-1}{2}]}\!\alpha_{p}\mathcal{L}_{\mathrm{ADM}}^{(p)}\,,\label{FLB_b}%
\end{align}
where the $p$th first-order Lagrangian density $\mathcal{L}^{(p)}_{\rm ADM}$ can be expressed, once the lapse and shift are reintroduced, as
\begin{eqnarray}
\mathcal{L}^{(p)}_{\rm ADM}&=&-\,\mathcal{Q}^{(p)}%
+2p\epsilon N\sqrt{|h|}\int_{0}%
^{1}\!ds\,\delta_{\lbrack j_{1}\cdots j_{2p}]}^{[i_{1}\cdots i_{2p}]}K_{i_{1}%
}^{j_{1}}K_{i_{2}}^{j_{2}}\left(  \frac{1}{2}R_{i_{3}i_{4}}^{j_{3}j_{4}}+\big(1-s^{2}\big)\epsilon
K_{i_{3}}^{j_{3}}K_{i_{4}}^{j_{4}}\right) \times \notag \\
&&\hspace{135pt}\cdots \times\left(  \frac
{1}{2}R_{i_{2p-1}i_{2p}}^{j_{2p-1}j_{2p}}+\big(1-s^{2}\big)\epsilon K_{i_{2p-1}%
}^{j_{2p-1}}K_{i_{2p}}^{j_{2p}}\right)\,, \label{LR} \\ 
&=&N\bar{\mathcal{L}}^{(p)}+2p\epsilon N\sqrt{|h|}\int_{0}^{1}\!ds(1-s)\delta_{\lbrack
j_{1}\cdots j_{2p}]}^{[i_{1}\cdots i_{2p}]}K_{i_{1}}^{j_{1}} K_{i_{2}}^{j_{2}}\left(  \frac{1}{2}\bar{R}_{i_{3}i_{4}}^{j_{3}j_{4}}-s^{2}\epsilon K_{i_{3}}^{j_{3}}K_{i_{4}}^{j_{4}}\right) \times%
\notag \\ 
&&\hspace{165pt}\cdots\times\left(  \frac{1}{2}\bar{R}_{i_{2p-1}i_{2p}}^{j_{2p-1}j_{2p}}%
-s^{2}\epsilon K_{i_{2p-1}}^{j_{2p-1}}K_{i_{2p}}^{j_{2p}}\right)\,,\label{LRbarR}
\end{eqnarray}
which explicitely eliminates second-order normal derivatives of $h_{ij}$ and where the second equality coincides with (\ref{LpINT}), thus confirming that the intrinsic function $r$ entering it is $\bar{\mathcal L}^{(p)}$.

This shows that, just as in the GR case, the Dirichlet action is equivalent to the
first-order action when we express all quantities in terms of $h_{ij}$, $K_{ij}$ and
$\bar{R}_{ijkl}$. Thus, $\mathcal L_{\rm ADM}=\sum \alpha_p\mathcal L_{\rm ADM}^{(p)}$ represents the first-order Lagrangian density for a generic Lovelock gravity theory.

In Ref. \cite{TZ} the authors obtain the expression
\begin{equation} \label{L1}
\pounds^{(p)}=N\sqrt{|h|}\sum_{i=0}^p %
\tilde{C}_{i(p)}
\delta_{[j_{1}\cdots j_{2p}]}^{[i_{1}\cdots i_{2p}]}%
R^{j_1j_2}_{i_1i_2}\cdots R^{j_{2i-1}j_{2i}}_{i_{2i-1}i_{2i}}%
K^{j_{2i+1}}_{j_{2i+1}}\cdots K^{j_{2p}}_{j_{2p}}\,,
\end{equation}
with coefficients 
\begin{equation}
\tilde{C}_{i(p)}=\frac{(-4)^{p-i}}{2\,i![2(p-i)-1]!!}\,.
\end{equation}

In order to compare (\ref{L1}) to our result $\mathcal{L}_{\mathrm{ADM}}^{(p)}$, we schematically represent $x=R^{ij}_{kl}$ and $y=K^i_j$ in Eq. (\ref{LR}) to obtain
\begin{equation}
\mathcal{L}_{\mathrm{ADM}}^{(p)} =\frac{x^p}{2^{p}}+2p\epsilon\int_{0}^{1}\!ds\,y^2\left(\frac{1}{2}x+\big(1-s^{2}\big)\epsilon y^2\right)^{p-1}=\sum_{i=0}^{p}C_{i(p)}x^iy^{2p-2i}\,,
\end{equation}
or, equivalently,
\begin{equation} \label{L0}
\mathcal{L}_{\mathrm{ADM}}^{(p)}=N\sqrt{|h|} \sum_{i=0}^{p}C_{i(p)}
\delta_{[j_{1}\cdots j_{2p}]}^{[i_{1}\cdots i_{2p}]}%
R^{j_1j_2}_{i_1i_2}\cdots R^{j_{2i-1}j_{2i}}_{i_{2i-1}i_{2i}}%
K^{j_{2i+1}}_{j_{2i+1}}\cdots K^{j_{2p}}_{j_{2p}}\,,
\end{equation}
where
\begin{equation}
C_{i(p)}=\frac{p!2^{p-2i}\epsilon^{p-i}}{i!(2(p-i)-1)!!}\,.
\end{equation}
Comparison between $\pounds^{(p)}$ and $\mathcal{L}_{\mathrm{ADM}}^{(p)}$ exhibits agreement up to an overall factor $p!/2^{p-1}$ due to different conventions.

Obtaining the Lovelock first-order Lagrangian densities $\mathcal L^{(p)}_{\rm ADM}$ through two straightforward routes, together with their explicit expressions in terms of $K_{ij}$ and $\bar R_{ijkl}$, see (\ref{LRbarR}), are the core results of the paper.\\

\textbf{The Gauss-Bonnet action.} As an example, consider the Gauss-Bonnet (GB) action supplemented with the Myers boundary term \cite{Myers,MuHo,Davis},
\begin{equation}\label{ID2}
I_{\rm Dir}[g]=\int\limits_{\mathcal{M}}d^Dx\,\mathcal{L}%
^{(2)}-\int\limits_{\partial\mathcal{M}}d^dx\,\beta^{(2)}\,,
\end{equation}
setting $\alpha_2=1$ for simplicity, where
\begin{equation}
\mathcal{L}^{(2)}=\sqrt{-g}\bigg(R^{\mu\nu\rho\sigma}R_{\mu\nu\rho\sigma}-%
4R^{\mu\nu}R_{\mu\nu}+R^2\bigg)=\sqrt{-g}R^{\mu\nu\rho\sigma}%
P_{\mu\nu\rho\sigma}
\end{equation}
is the Gauss-Bonnet scalar density, and where
\begin{align}
  P^{\mu\nu}_{\rho\sigma}&=\frac{1}{4}\delta^{[\mu\nu\alpha_1\alpha_2]}_{[\rho\sigma\beta_1\beta_{2}]}R_{\alpha_1\alpha_2}^{\beta_1\beta_{2}}\nonumber\\
  &=R^{\mu\nu}_{\rho\sigma}-2\delta^\mu_{[\rho}R_{\sigma]}^\nu+2\delta^\nu_{[\rho}R_{\sigma]}^\mu+\delta^\mu_{[\rho}\delta_{\sigma]}^\nu R
\end{align}
has the symmetries of the Riemann tensor and is divergenceless ($\nabla_\mu P^{\mu}_{\ \, \nu\rho\sigma}=0$) due to the Bianchi identities. Here brackets denote antisymmetrization, as in $A^\mu_{[\rho}B_{\sigma]}^\nu=\frac{1}{2}(A^\mu_{\rho}B_{\sigma}^\nu-A^\mu_{\sigma}B_{\rho}^\nu)$. Finally,
\begin{eqnarray}
\beta^{(2)}=-2\epsilon\sqrt{|h|}\delta_{[j_1j_2j_3]}^{[i_1i_2i_3]}K^{j_1}_{i_1}\left(\bar{R}^{j_2j_3}_{i_2i_3}-\frac{2\epsilon}{3}%
K^{j_2}_{i_2}K^{j_3}_{i_3} \right)=-4\epsilon \left(J-2\,%
\bar G^i_jK^j_i\right)\, ,\label{boundaryTermCovariantEGB}\\
\text{with}\quad \epsilon J^i_j=\frac{1}{3}K^i_j\!\left(K^k_lK^l_k-K^2\right)+\frac{2}{3}K K^i_kK^k_j-\frac{2}{3}K^i_kK^k_lK^l_j\quad\text{and}\quad J=J^k_k\,. \notag
\end{eqnarray}
This case has been studied  in, e.g, Refs. \cite{MisOl,DM,DMO} and generalized to Einstein-scalar-Gauss-Bonnet theories in \cite{Julie:2020vov}.
In Gaussian coordinates, the variation of (\ref{ID2}) adopts the form
\begin{equation}
\delta I_{\rm Dir}=\int\limits_{\mathcal M}d^Dx\sqrt{-g}H^{\mu\nu}\delta g_{\mu\nu}%
+\!\int\limits_{\partial \mathcal{M}}d^d x\sqrt{|h|}\pi_{(2)}^{ij}\delta h_{ij}\ ,\label{adaptedVariationEGB}
\end{equation}
where
\begin{align}
H^\mu_\nu=-\frac{1}{8}\delta^{[\mu\,\mu_1\mu_2\mu_3\mu_4]}_{[\,\nu\,\nu_1\nu_2\nu_3\nu_4]}R_{\mu_1\mu_2}^{\nu_1\nu_2}R_{\mu_3\mu_4}^{\nu_3\nu_4}
\end{align}
is the Lanczos tensor and where
\begin{eqnarray}
\pi_{(2)}^{ij}&=&\epsilon\sqrt{|h|}h^{ik}\delta^{[jj_1j_2j_{3}]}_{[ki_1i_2 i_{3}]}%
K^{i_1}_{j_1}\Bigg(\bar R^{i_2i_3}_{j_2j_3}-\frac{2\epsilon}{3}K^{i_2}_{j_2}K^{i_3}_{j_3}\Bigg) \notag \\
&=& 2\epsilon\sqrt{|h|}\bigg(2h^{mj}\bar{P}^{ik}_{ml}K^l_k-3J^{ij}+h^{ij}J\big)\bigg)\,.\label{Pi2}
\end{eqnarray}

The tensor density (\ref{Pi2}) is the canonical momentum associated to the first-order action. Hence, solving
\begin{align}
\frac{\partial \mathcal L^{(2)}_{\rm ADM}}{\partial(\partial_wh_{ij})}&=\pi^{ij}_{(2)}\,,\label{eqDiffLagrangianEGB}
\end{align}
we find (after inclusion of the lapse and shift) 
\begin{align}
\mathcal L^{(2)}_{\rm ADM}&=N\bar{\mathcal L}^{(2)}+\,N\sqrt{|h|} \delta^{[i_1i_2i_3i_4]}_{[j_1j_2j_3j_4]}\left[\epsilon K^{j_1}_{i_1}K^{j_2}_{i_2}\left( \bar R^{j_3j_4}_{i_3i_4}-\frac{\epsilon}{3}K^{j_3}_{i_3}K^{j_4}_{i_4}\right)\right]\nonumber\\
&=N\bar{\mathcal L}^{(2)}+N\sqrt{|h|}\left[4\epsilon\bar P^{ij}_{kl}K^k_iK^l_j+KJ-3K^i_jJ^j_i\right]\,, \label{lagrangianADM}
\end{align}
where the first term is obtained by identifying it to the restriction of the Gauss-Bonnet Lagrangian density to the surface $w=cst$, that is building it with the intrinsic curvature only:
\begin{equation}
\bar{\mathcal L}^{(2)}=\frac{1}{4}\sqrt{|h|}\,\delta^{[i_1i_2i_3i_4]}_{[j_1j_2j_3j_4]}\bar R_{i_1i_2}^{j_1j_2}\bar R_{i_3i_4}^{j_3j_4}\,.
\end{equation}
When $D=4$ (i.e. $d=3$), the Lanczos tensor, the momentum and the generalized ADM Lagangian vanish, as evident from their expression in terms of rank-five and rank-four Kronecker deltas, respectively.

On the other hand, in Appendix \ref{ED}, the decomposition of $\mathcal{L}^{(2)}$ shows that the same Lagrangian density can be obtained by bulkanization. Using Eq. (\ref{ET2}), the Lagrangian density in Eq. (\ref{ID2}) can be shown to yield the same result, that is (\ref{lagrangianADM}).

\section{Hamiltonian Dynamics}

In order to define an ordinary Hamiltonian, a first-order Lagrangian density $\mathcal L_{\rm ADM}$ is required. If the induced metric $h_{ij}$ is chosen as the dynamical variable, the Hamiltonian is given by the Legendre transformation 
\begin{eqnarray}\label{HadmK}
H=\int d^dx\big(\pi^{ij}\partial_wh_{ij}-\mathcal{L}%
_{\rm ADM}\big)\,,
\end{eqnarray}
where the canonical momentum $\pi^{ij}$ is defined as
\begin{equation}\label{pidef}
\pi^{ij}\equiv\frac{\partial \mathcal L_{\rm ADM}}{\partial(\partial_w h_{ij})}\,.
\end{equation}
This functional must be written in terms of $h_{ij}$ and $\pi^{ij}$. This is the path chosen by Arnowitt, Deser and Misner to construct their celebrated Hamiltonian.

The same path can be taken to construct a Hamiltonian from the first-order Lagrangian density of Lovelock gravity found in the previous section. For each $p$-th contribution, the associated Hamiltonian is computed as
\begin{eqnarray} \label{HLov}
H^{(p)}=\int d^dx\big(\pi^{ij}_{(p)}\partial_wh_{ij}-\mathcal{L}%
_{\rm ADM}^{(p)}\big)\,.
\end{eqnarray}

From the canonical momentum (\ref{canmom}), and in Gaussian coordinates, we have
\begin{eqnarray}
\pi^{ij}_{(p)}\partial_wh_{ij}=2NK^i_j\pi^j_{(p)i}&=&2p\epsilon N\sqrt{|h|}\int_{0}%
^{1}\!ds\,\delta_{\lbrack j_{1}\cdots j_{2p}]}^{[i_{1}\cdots i_{2p}]}K_{i_{1}%
}^{j_{1}}K_{i_{2}}^{j_{2}}\Big( \frac{1}{2}\bar{R}_{i_{3}i_{4}}^{j_{3}j_{4}} -s^{2}\epsilon K_{i_{3}}^{j_{3}}K_{i_{4}}^{j_{4}}\Big)  \times\cdots \notag \\
&&\hspace{90pt}\cdots\times\left( \frac
{1}{2}\bar{R}_{i_{2p-1}i_{2p}}^{j_{2p-1}j_{2p}}-s^{2}\epsilon K_{i_{2p-1}%
}^{j_{2p-1}}K_{i_{2p}}^{j_{2p}}\right)\,,\quad
\end{eqnarray}
which identifies to the last term of the second member of Eq. (\ref{MET}). Therefore the $p$-th Hamiltonian density $\mathscr{H}^{(p)}$ identifies, in the Gaussian gauge (\ref{mmetric}), to the functional $\mathcal{Q}^{(p)}$ (which is proportionnal to $\mathcal E^w_{(p)w}$, as mentioned below (\ref{Hp})).
The lapse and shift $N^i$ can then be restored using Eq. (\ref{Kadm}) to find the full Hamiltonian:
\begin{equation}
H=\int d^{d}x\,\bigg(N\mathscr{H}+N^{i}%
\mathscr{H}_{i}\bigg)\,,\label{HG}%
\end{equation}
where the Hamiltonian constraints take the form 
\begin{eqnarray}\label{EOMLL1}
\mathscr{H}&=&\sum_{p=0}^{[\frac{D-1}{2}]}\alpha_{p}\,\mathscr{H}^{(p)}\,, \notag \\
\mathscr{H}_{i}&=&-2\bar{\nabla}_{j}\pi_{i}^{j}\,,
\end{eqnarray}
 where $\mathscr{H}^{(p)}=\mathcal{Q}^{(p)}/N$ and $\pi^i_j$ are given respectively in Eqs. (\ref{Hp}) and (\ref{canmom}).

Due to the non linear relation between $\pi^{ij}$ and $K_{ij}$, it is not possible in general to write $K_{ij}$ in terms of $\pi^{ij}$. Thus, the Hamiltonian above is only given implicitly in terms of the momenta. 
On the other hand, it is an exercise to check that the components of the Lovelock tensor $\mathcal E^\mu_\nu$ defined in (\ref{eom}) verify $\mathcal{E}_{w}^{w}=\mathscr{H}/2\sqrt
{|h|}$ and $\mathcal{E}^w_{i}=\mathscr{H}_{i}/2N\sqrt{|h|}$ in Gaussian coordinates, while ${\mathcal E^i_{(p)j}}$ reads
\begin{eqnarray} \label{EOMij}
\mathcal{E}^i_{(p)j}&=&-p\epsilon\int_0^1\!ds\,\delta^{[ii_1\cdots i_{2p}]}_{[jj_1\cdots j_{2p}]}%
K^{j_1}_{i_1}K^{j_2}_{i_2}\left( \frac{1}{2}\bar R^{j_3j_4}_{i_3i_4}-\epsilon s^2K^{j_3}_{i_3}%
K^{j_4}_{i_4} \right)\times \cdots\times \Bigg( \frac{1}{2}\bar R^{j_{2p-1}j_{2p}}_{i_{2p-1}i_{2p}} \notag \\
&& -\epsilon s^2K^{j_{2p-1}}_{i_{2p-1}}K^{j_{2p}}_{i_{2p}}\Bigg)-p\epsilon\int_0^1\!ds\,\delta^{[ii_2\cdots i_{2p}]}_{[j_1\cdots j_{2p}]}%
K^{j_1}_{j}K^{j_2}_{i_2}\left( \frac{1}{2}\bar R^{j_3j_4}_{i_3i_4}-\epsilon s^2K^{j_3}_{i_3}%
K^{j_4}_{j_4} \right)\times \cdots \notag \\
&& \cdots \times \Bigg( \frac{1}{2}\bar R^{j_{2p-1}j_{2p}}_{i_{2p-1}i_{2p}}-%
\epsilon s^2K^{j_{2p-1}}_{i_{2p-1}}K^{j_{2p}}_{i_{2p}}\Bigg)-\frac{1}{2^{p+1}}\delta^{[ii_1\cdots i_{2p}]}_{[jj_1\cdots j_{2p}]}R^{j_1j_2}_{i_1i_2}\times\cdots \times R^{j_{2p-1}j_{2p}}_{i_{2p-1}i_{2p}}\notag \\
&&-\frac{1}{2}\delta^{[ii_1\cdots i_{2p-1}]}_{[jj_1\cdots j_{2p-1}]}\bar{\nabla}_{i_1}\bigg(K^{j_2}_{i_2}\bar{\nabla}^{j_1}K^{j_3}_{i_3}R^{j_4j_5}_{i_4i_5}\times\cdots\times R^{j_{2p-2}j_{2p-1}}_{i_{2p-2}i_{2p-1}}\bigg)+\frac{\partial_w\big(\pi^i_j\big)}{N\sqrt{|h|}}\quad
\end{eqnarray}
where $R_{ijkl}$ is understood as an implicit function of $\bar R_{ijkl}$ and $K_{ij}$; see Eq. (\ref{EijL}) for completeness. Here we gathered terms which are equal to the normal derivative of $\pi^i_j$ using the tools presented in Appendix \ref{ED} 
(for its explicit expansion in the scalar-Gauss-Bonnet case, see \cite{Julie:2020vov}).

The Lagrangian and Hamiltonian dynamics are equivalent and the correspondence between the field equations is given by
\begin{eqnarray} \label{EOMEq}
\frac{\delta H}{\delta N}=0 & \Leftrightarrow & \mathcal{E}^w_w=0 \,, \notag \\
\frac{\delta H}{\delta N^i}=0 & \Leftrightarrow & \mathcal{E}^w_i=0 \,. 
\end{eqnarray}
In addition, by definition of $H$ we have that 
\begin{equation}
\frac{\delta H}{\delta h_{ij}}\Bigg|_{\pi^{ij}}=-\frac{\delta L}{\delta h_{ij}}\Bigg|_{\partial_w h_{ij}}\quad {\rm where}\quad L=\int d^dx\,\mathcal{L}_{\rm ADM}\,.
\end{equation}
Hence, it can be checked explicitly using the equation above and (\ref{LRbarR}) that
\begin{equation}
\frac{\delta H}{\delta h_{ij}}\Bigg|_{\pi^{ij}}=-\partial_w\pi^{ij} \, \Leftrightarrow \, \mathcal{E}^{ij}=0 \,.
\end{equation}

In the case of GR, we also have that $\frac{\delta H}{\delta \pi^{ij}}=\partial_wh_{ij}  \Leftrightarrow  K_{ij}=\frac{1}{2N}\partial_wh_{ij}$. This relation cannot be proven in the general Lovelock case, as it requires the invertibility of $\pi^{ij}$. However, it does not provide provide extra dynamical information. 

The particular case of Gauss-Bonnet gives 
\begin{eqnarray}
H^{(2)}&=&-\int d^dx\,N\bar{\mathcal{L}}^{(2)}+\epsilon\int d^dx\,%
N\sqrt{|h|}\delta^{[i_1i_2i_3i_4]}_{[j_1j_2j_3j_4]}%
K^{j_1}_{i_1}K^{j_2}_{i_2}\Bigg(\bar R^{j_3j_4}_{i_3i_4}-%
 \epsilon K^{j_3}_{i_3}K^{j_4}_{i_4}\Bigg)\notag \\
&=&-\int d^dx\,N\bar{\mathcal{L}}^{(2)}+\int d^dx\,N\sqrt{|h|}%
\Big(2\epsilon\bar{P}_{ijkl}K^{ik}K^{jl}-\frac{1}{2}K^4+3K^2K_{j}^{i}K_{i}^{j} \notag \\
&&\hspace{100pt}-4KK_{j}^{i}K_{k}^{j}%
K_{i}^{k}-\frac{3}{2}K_{j}^{i}K_{i}^{j}K_{l}^{k}K_{k}^{l}+3K_{j}^{i}K_{k}^{j}K_{l}^{k}K_{i}^{l}\Big)\,,\quad
\end{eqnarray}
where in the second line we have just expanded  the generalized Kronecker delta.

\section*{Conclusions}

In this paper we investigated the links between the Dirichlet variational principle, and the first-order Lagrangian density and Hamiltonian of Lovelock gravity. 
Starting from the simple example of a Lagrangian linear in the acceleration in point mechanics, we have identified two methods to compute the associated first-order Lagrangian: integration of the momentum and bulkanization of boundary terms. 
We then worked out the case of General Relativity to recover the ADM Lagrangian density from the Dirichlet action.

More powerful, however, is the use of the momentum integration and bulkanization methods to obtain the first-order Lagrangian density of Lovelock gravity. Bulkanizing the Myers term explicitly eliminates all second-order normal derivatives in the bulk. In Gaussian coordinates, the resulting Lagrangian density has the form $\mathcal{L}^{(p)}_{\rm ADM}=\pi^{ij}\partial_wh_{ij}-2N\sqrt{|h|}\mathcal E^w_w$, making manifest the connection with the Hamiltonian formalism. Indeed, a Legendre transformation of the first-order Lagrangian density, directly gives the Hamiltonian density of the system $N\mathscr{H}^{(p)}=2N\sqrt{|h|}\mathcal{E}^w_w$. In addition, we have that the Lagrangian and Hamiltonian formalisms are equivalent at the level of the dynamics and surface terms. Indeed, the variation of the Hamiltonian action 
\begin{equation}
I_{H}=\int\limits_{\mathcal{M}}d^Dx\,\bigg( \pi^{ij}\partial_wh_{ij}-\mathcal{L}_{\rm ADM}\bigg)\,,
\end{equation} 
produces --on-shell--
\begin{equation}
\delta I_{H}=\int\limits_{\partial \mathcal{M}}d^dx\,\pi^{ij}\delta h_{ij}\,.
\end{equation}
This matches the surface term obtained in Eq. (\ref{ILV}) from the variation of the first-order Lagrangian. This fact will be employed in future work to define junction conditions for thin shells \`a la Hamilton for Lovelock gravity. 

Our methods should also be useful to generalize the Arnowitt-Deser-Misner (ADM) mass formula to Lovelock gravities. In fact, the canonical momentum readily defines a conserved current when contracted with a boundary Killing vector.

For an arbitrary set of couplings in the Lovelock action, some of the components of the metric solution may not be fully determined by the field equations \cite{Whee}. For instance, the component $g_{tt}$ of any static spherically symmetric ansatz remains arbitrary if the action has non-unique degenerate vacuum. This problem can be avoided by a given choice of the coefficients (e.g., the cases of GR, Chern-Simons, Born-Infeld and Pure Lovelock \cite{Cai:2006pq,Dadhich:2010gu,Aranguiz:2015voa}). However, the higher curvature terms in the action make the symplectic matrix change the rank for certain backgrounds, generating extra local symmetries and decreasing degrees of freedom in some sectors of the space of solutions \cite{BaGaHe,BaGaHe2,DaDuMeMi,GiMeMiZa}. 
This kind of degeneracy in Lovelock gravity also occurs in cosmological solutions \cite{DerFar}, where the field equations cannot predict the evolution of the scale factor $a(t)$ because the coefficient of $\ddot a(t)$ goes through zero during the evolution. This also renders the hamiltonian quantization of the system problematic \cite{TZ}.


\section*{Acknowledgements}
We warmly thank Nathalie Deruelle for very stimulating discussions and for insightful suggestions along the preparation of this work. P.G. is a UNAB Ph.D. Scholarship holder. F.L.J. is supported by NSF Grants No. PHY-1912550 and AST-2006538, NASA ATP Grants No. 17-ATP17-0225 and 19-ATP19-0051, and NSF-XSEDE Grant No. PHY-090003. This work has received funding from the European Union’s Horizon 2020 research and innovation programme under the Marie Skłodowska-Curie grant agreement No. 690904 and networking support by the GWverse COST Action CA16104, ``Black holes, gravitational waves and fundamental physics". This work was also funded by FONDECYT grant 11180894 "Covariant boundary terms and conservation laws in modified gravity" (N.M.) and FONDECYT grant 1170765 “Boundary dynamics in anti-de Sitter gravity and gauge/gravity duality" (R.O.).

\section*{Appendices}
\appendix
\section{Conventions} \label{GC}

In this paper we set $16\pi G=c=1$. Throughout the text $g$ is the determinant of the metric $g_{\mu\nu}$ (with inverse $g^{\mu\nu}$), $R^\mu_{\ \nu\rho\sigma}\equiv\partial_\rho\Gamma^\mu_{\nu\sigma}-\cdots$ is the Riemann tensor where $\Gamma^\mu_{\nu\sigma}\equiv{\scriptstyle{1\over2}}g^{\mu\lambda}(\partial_\nu g_{\sigma\lambda}+\cdots)$ are the Christoffel symbols, $R_{\mu\nu}\equiv R^\lambda_{\ \mu\lambda\nu}$ is the Ricci tensor and $R\equiv g^{\mu\nu}R_{\mu\nu}$ is the scalar curvature.

In Gaussian coordinates
\begin{equation}
ds^{2}=\epsilon\,N^2(w)dw^{2}+h_{ij}\left(  w,x^{i}\right)
dx^{i}dx^{j}\,,\label{GaussianCoordAppendix}
\end{equation}
 the non-vanishing components of the Christoffel symbols are $\Gamma^i_{jk}=\bar\Gamma^i_{jk}(h)$ and
\begin{equation}
\Gamma_{ij}^{w}=-\frac{\epsilon}{2N^2}\partial_wh_{ij}\ ,\ \ \ \Gamma
_{wj}^{i}=\frac{1}{2}h^{ik}\partial_wh_{jk}\,\quad %
\Gamma_{ww}^{w}=\frac{\partial_w N}{N} . \label{gen_gamma}%
\end{equation}
The normal to a surface $\Sigma_w$ of constant $w$ is defined as
\begin{equation}
n_{\mu}=\epsilon N\delta_{\mu}^{w}\,, \label{normal_choice}%
\end{equation}
so that $n_{\mu}n^{\mu}=\epsilon$. On the other hand, the
extrinsic curvature is defined as
\[
K_{ij}=V_{i}^{\mu}V_{j}^{\nu}\nabla_{\mu}n_{\nu}\,,%
\]
where $V_{i}^{\mu}$ are the projectors on
the corresponding surface. In Gaussian coordinates $V_{i}^{\mu}=\delta_{i}^{\mu}$ and as a
consequence of the normal vector definition (\ref{normal_choice}), the extrinsic curvature  is given in terms of $h_{ij}$
by
\begin{equation}
K_{ij}=\nabla_{i}n_{j}=-\epsilon N\Gamma_{ij}^{w}=\frac{1}{2N}\partial_{w}h_{ij}\,. \label{K_choice}%
\end{equation}
Consequently, the Christoffel symbols satisfy%
\begin{equation} \label{chrGN}
\Gamma^i_{wj}=NK^i_j \quad , \quad \Gamma^w_{ij}=-\frac{\epsilon}{N} K_{ij}\,,
\end{equation}
and the curvature tensors have the form
\begin{eqnarray} 
R^{ij}_{kl}&=&\bar R^{ij}_{kl}-\epsilon\big (K^i_kK^j_l-K^i_lK^j_k\big)\,, \label{Rijkl} \\
R^{wi}_{jk}&=&-\frac{\epsilon}{N}\left(\bar{\nabla}_jK^i_k-%
\bar{\nabla}_kK^i_j\right)\,, \label{wijk}\\
R^{ij}_{wk}&=&-  N\left (\bar{\nabla}^iK^j_k-\bar{\nabla}^jK^i_k\right )\,, \label{Rijwk} \\
R^{wi}_{wj}&=&-\frac{\epsilon}{N}\,\partial_wK^i_j-\epsilon K^i_k K^k_j\,,\label{Rwiwj}\\
R^i_j&=&\bar R^i_j-\epsilon KK^i_j-\frac{\epsilon}{N}\,\partial_wK^i_j\,,\label{Rij}\\
R^w_i&=&-\frac{\epsilon}{N}\bar{\nabla}_j\big(K\delta^j_i-K^j_i\big)\,,\label{Rwi} \\
R^w_w&=&-\frac{\epsilon}{N}\,\partial_wK-\epsilon K^i_jK^j_i\,,\label{Rww} \\
R&=&\bar R-\epsilon\big(K^2+K^i_jK^j_i\big)-\frac{2\epsilon}{N}\,\partial_wK\,. \label{Rscalar}
\end{eqnarray}
The equations above are the Gauss-Codazzi-Mainardi relations in tensorial language and in Gaussian coordinates.

\section{Bulkanization of Myers terms} \label{ED}
As a warmup exercise, let us consider the integral of $\beta^{(2)}$, see (\ref{boundaryTermCovariantEGB}), on the boundary $\partial\mathcal M=\Sigma_{w_i}\cup\Sigma_{w_f}\cup\mathcal C$, which is the union of the surfaces $w=w_i$ and $w=w_f$ and of their complement $\mathcal C$. Its bulkanization yields:
\begin{eqnarray}\label{SP}
\int\limits_{\partial \mathcal{M}}d^dx\, \beta^{(2)}=-2\epsilon\int\limits_{\mathcal{M}} d^Dx\,\partial_w\left [\sqrt{|h|}\delta^{[j_1j_2 j_{3}]}_{[i_1i_2 i_{3}]}K^{i_1}_{j_1}\left(\bar R^{i_2i_3}_{j_2j_3}-\frac{2\epsilon}{3}K^{i_2}_{j_2}K^{i_3}_{j_3} \right)\right ]\,,
\end{eqnarray}
 modulo a contribution on $\mathcal C$ which can be discarded for our purposes, see below (\ref{a2}).

In order to compute the normal derivatives
involved and construct the desired structures, it is useful to rewrite $\partial_wK^{i_1}_{j_1}$ using (\ref{Rwiwj}) as
\begin{equation} \label{Kp}
\partial_wK^{i_1}_{j_1}=-\epsilon N\left (R^{wi_1}_{wj_1}+\epsilon %
K^{i_1}_lK^l_{j_1}\right )\,,
\end{equation}
and
\begin{equation} \label{hp}
\partial_w\sqrt{|h|}=\frac{1}{2}\sqrt{|h|}h^{ij}\partial_wh_{ij}= NK\sqrt{|h|}\,,
\end{equation}
where $K$ is the trace of the extrinsic curvature. Since moreover
$
\partial_w\bar R^i_{\,jkl}=\bar\nabla_k(\partial_w\bar\Gamma^i_{jl})-\bar\nabla_l(\partial_w\bar\Gamma^i_{jk})
$ with
$
\partial_w\bar{\Gamma}^{k}_{ij}=N\big(\bar{\nabla}_iK^k_j+\bar{\nabla}_jK^k_i-\bar{\nabla}^kK_{ij}\big)
$
(which exhibits $\partial_w\bar{\Gamma}^{k}_{ij}$ as an intrinsic tensor),
a short calculation yields
\begin{equation}
\delta_{[i_{1}i_{2}i_{3}]}^{[j_{1}j_{2}j_{3}]}K_{j_{1}}^{i_{1}}\partial_w
\bar R_{j_{2}j_{3}}^{i_{2}i_{3}} =-2N \delta_{[i_{1}i_{2}i_{3}]}^{[j_{1}j_{2}j_{3}]}K_{j_{1}}^{i_{1}}\left[
K_{k}^{i_{3}}R_{\ j_{2}j_{3}}^{i_{2}k}+2\epsilon K_{k}^{i_{3}}K_{j_{2}}^{i_{2}}%
K_{j_{3}}^{k}+2\bar{\nabla}_{j_{2}}\bar{\nabla}^{i_{2}}K_{j_{3}}^{i_{3}%
}\right] \,. \label{R_prime}%
\end{equation}
Combining the results above, (\ref{SP}) can be rewritten as
\begin{eqnarray}
&&\int\limits_{\partial \mathcal{M}}d^dx\,\beta^{(2)}=-2\epsilon\int\limits_{\mathcal{M}} d^Dx\,N\sqrt{|h|}\delta^{[j_1j_2j_{3}]}_{[i_1i_2i_{3}]}\bigg(\!\!-2 K^{i_1}_{j_1}K^{i_2}_kR^{ki_3}_{j_2j_3}-%
4 \epsilon K^{i_1}_{j_1}K^{i_2}_{j_2}K^{i_3}_{k}K^{k}_{j_3}\hspace{60pt} \notag\\
&&\quad -4 K^{i_1}_{j_1}\bar{\nabla}_{j_2}\bar{\nabla}^{i_2}K^{i_3}_{j_3} -\epsilon \left (R^{wi_1}_{wj_1}+\epsilon%
K^{i_1}_lK^l_{j_1}\right )R^{i_2i_3}_{j_2j_3}+K^{i_1}_{j_1}\left(R^{i_2i_3}_{j_2j_3}+%
\frac{4\epsilon}{3}K^{i_2}_{j_2}K^{i_3}_{j_3} \right)K  \bigg )\,.
\label{almost}
\end{eqnarray}
At this point, we can use the identities  (\ref{KKKK}) and (\ref{KKR}) to find
\begin{eqnarray} \label{amst}
\int\limits_{\partial \mathcal{M}}d^dx\,\beta^{(2)}&=&2\int\limits_{\mathcal{M}} d^Dx\,N\sqrt{|h|}\delta^{[j_1j_2j_{3}]}_{[i_1i_2i_{3}]}\left (R^{wi_1}_{wj_1}R^{i_2i_3}_{j_2j_3}+4\epsilon K^{i_1}_{j_1}\bar{\nabla}_{j_2}\bar{\nabla}^{i_2}K^{i_3}_{j_3}\right )  \notag \\
&&\qquad -2\epsilon\int\limits_{\mathcal{M}} d^Dx\,N\sqrt{|h|}\delta^{[j_1j_2j_{3}j_4]}_{[i_1i_2i_{3}i_4]}K^{i_1}_{j_1}K^{i_2}_{j_2}\left(\bar R^{i_3i_4}_{j_3j_4}-\frac{2\epsilon}{3}K^{i_3}_{j_3}K^{i_4}_{j_4} \right)\,.
\end{eqnarray}
We notice that $R^{wi_1}_{wj_1}$ in the first term contains normal derivatives of the extrinsic curvature, see (\ref{Rwiwj}), that will cancel out with those coming from the expanded Gauss-Bonnet Lagrangian density,
\begin{eqnarray}
\mathcal{L}^{(2)}&=&2N\sqrt{|h|}%
\delta^{[i_1i_2 i_3]}_{[j_1j_2 j_3]}\Big (R^{wj_1}_{wi_1}R^{j_2j_3}_{i_2i_3}+%
R^{wj_1}_{i_1i_2}R^{j_2j_3}_{wi_3}\big )
+\frac{1}{4}N\sqrt{|h|}%
\delta^{[i_1\cdots i_4]}_{[j_1\cdots j_4]}R^{j_1j_2}_{i_1i_2}R^{j_3j_4}_{i_3i_4}\,. \hspace{10pt}
\end{eqnarray}
Using $\delta^{[j_1j_2j_{3}]}_{[i_1i_2i_{3}]}R^{wi_1}_{j_1j_2}R^{i_2i_3}_{wj_3}=-4\epsilon\delta^{[j_1j_2j_{3}]}_{[i_1i_2i_{3}]}\bar{\nabla}_{j_2}K^{i_1}_{j_1}\bar{\nabla}^{i_2}K^{i_3}_{j_3}$ and integrating by parts we get
\begin{eqnarray} \label{GBFOL}
\int\limits_\mathcal{M} d^Dx\,\mathcal{L}^{(2)}&=&2\int\limits_{\mathcal{M}} d^Dx\,N\sqrt{|h|}\delta^{[j_1j_2j_{3}]}_{[i_1i_2i_{3}]}\Big(R^{wi_1}_{wj_1}R^{i_2i_3}_{j_2j_3}+4\epsilon\bar{\nabla}^{i_2}K^{i_1}_{j_1}\bar{\nabla}_{j_2}K^{i_3}_{j_3}\Big )+ \notag \\
&&\hspace{130pt}+\frac{1}{4}\int\limits_\mathcal{M} d^Dx\,N\sqrt{|h|}%
\delta^{[i_1\cdots i_4]}_{[j_1\cdots j_4]}R^{j_1j_2}_{i_1i_2}%
R^{j_3j_4}_{i_3i_4}\,, \hspace{10pt}
\end{eqnarray}
where we discarded terms that are total $\bar\nabla_i$ derivatives, i.e. terms living on $\mathcal C$. 

Subtracting (\ref{GBFOL}) and (\ref{amst}) we finally get
\begin{eqnarray}\label{ET2}
 \int\limits_\mathcal{M} d^{D}x\,\Big(\mathcal{L}^{(2)}-\frac{d}{dw}\big(\beta^{(2)}\big)\Big)&=&-\int\limits_\mathcal{M}d^{D}x\, %
 \mathcal{Q}^{(2)}+2\epsilon\int\limits_\mathcal{M}d^{D}x\,N\sqrt{|h|}\delta^{[i_1\cdots %
 i_{4}]}_{[j_1\cdots j_{4}]}%
K^{j_1}_{i_1}K^{j_2}_{i_2}\times \notag \\
&&\hspace{110pt}\times\left(\bar R^{j_3j_4}_{i_3i_4}-%
\frac{2\epsilon}{3}K^{j_3}_{i_3}K^{j_4}_{i_4} \right)\,.
\end{eqnarray}
where $\mathcal{Q}^{(2)}$ is obtained by setting $p=2$ in Eq. (\ref{Hp}).

The same bulkanization procedure can be performed for any Lovelock density with its corresponding Myers term.
The use of Eqs. (\ref{Kp}), (\ref{hp}), (\ref{R_prime}), (\ref{K2m}) and similar steps to those described above yield
\begin{eqnarray} \label{Bp}
\frac{d}{dw}\big(\beta^{(p)}\big)=-2p\epsilon N\sqrt{|h|}\int_0^1\!ds\,\delta^{[i_1\cdots i_{2p}]}_{[j_1\cdots j_{2p}]}%
K^{i_1}_{j_1}K^{i_2}_{j_2}\left( \frac{1}{2}\bar R^{i_3i_4}_{j_3j_4}-\epsilon s^2K^{i_3}_{j_3}%
K^{i_4}_{j_4} \right)\times \hspace{100pt}&&\notag \\
\hspace{100pt}\times \left( \frac{1}{2}\bar R^{i_{2p-1}i_{2p}}_{j_{2p-1}j_{2p}}-\epsilon s^2K^{i_{2p-1}}_{j_{2p-1}}%
K^{i_{2p}}_{j_{2p}} \right)+\frac{p}{2^{p-2}}\delta^{[i_1\cdots i_{2p-1}]}_{[j_1\cdots j_{2p-1}]}%
\Big(R^{wj_1}_{wi_2}R^{j_2j_3}_{i_2i_3}+\hspace{20pt}&&\notag \\
+(p-1)R^{wj_1}_{i_1i_2}R^{j_2j_3}_{wi_3}%
\Big )R^{j_4j_5}_{i_4i_5}\times\cdots\times R^{j_{2p-2}j_{2p-1}}_{i_{2p-2}i_{2p-1}}\,.&& \notag
\end{eqnarray}
Since the expanded Lagrangian density $\mathcal{L}^{(p)}$ takes the form
\begin{eqnarray} \label{FoL}
\mathcal{L}^{(p)}=\frac{p}{2^{p-2}}\sqrt{|h|}\delta^{[i_1\cdots i_{2p-1}]}_{[j_1\cdots j_{2p-1}]}\Big (%
R^{wj_1}_{wi_1}R^{j_2j_3}_{i_2i_3}+(p-1)R^{wj_1}_{i_1i_2}R^{j_2j_3}_{wi_3}\Big )R^{j_4j_5}_{i_4i_5}\times%
\cdots\times R^{j_{2p-2}j_{2p-1}}_{i_{2p-2}i_{2p-1}}\notag \\
+\frac{1}{2^p}N\sqrt{|h|}\delta^{[i_1\cdots i_{2p}]}_{[j_1\cdots j_{2p}]}R^{j_1j_2}_{i_1i_2}\times\cdots \times R^{j_{2p-1}j_{2p}}_{i_{2p-1}i_{2p}}\,, \hspace{15pt}
\end{eqnarray}
we get
\begin{eqnarray} \label{ABF}
 \int\limits_\mathcal{M}d^{D}x\,\Big( \mathcal{L}^{(p)}-\frac{d}{dw}\big(\beta^{(p)}\big)\Big)&=& -\int\limits_\mathcal{M}d^{D}x\, \mathcal{Q}^{(p)} +2p\epsilon\int\limits_\mathcal{M}d^{D}x\,N\sqrt{|h|}\int_0^1\!ds\,\delta^{[i_1\cdots i_{2p}]}_{[j_1\cdots j_{2p}]}%
K^{j_1}_{i_1}K^{j_2}_{i_2}\times \notag \\
&&\hspace{20pt}\times\left( \frac{1}{2}\bar R^{j_3j_4}_{i_3i_4}-\epsilon s^2K^{j_3}_{i_3}%
K^{j_4}_{i_4} \right)\times \cdots\times %
\Big( \frac{1}{2}\bar R^{j_{2p-1}j_{2p}}_{i_{2p-1}i_{2p}}\notag \\
&&\hspace{160pt}-\epsilon s^2K^{j_{2p-1}}_{i_{2p-1}}K^{j_{2p}}_{i_{2p}} \Big)\,.
\end{eqnarray}

Finally, the same game can be played when projecting the equations of motion 
$\mathcal{E}^i_j$: we can see that
\begin{equation}
\mathcal{E}^i_{(p)j}=-\frac{1}{2^{p+1}}\delta^{[i\mu_1\cdots\mu_{2p}]}%
_{[j\nu_1\cdots\nu_{2p}]}R^{\nu_1\nu_2}_{\mu_1\mu_2}\cdots R^{\mu_{2p-1}%
\mu_{2p}}_{\nu_{2p-1}\nu_{2p}}\,
\end{equation}
exhibits the same structure as $\mathcal{L}^{(p)}$ except for the extra pair of indices. According to Eq. (\ref{K2mij}), we will need an extra term 
when packing the terms in a one-rank-higher delta and get, restoring the lapse and shift,
\begin{eqnarray} 
\mathcal{E}^i_{(p)j}&=&-p\epsilon\int_0^1\!ds\,\delta^{[ii_1\cdots i_{2p}]}_{[jj_1\cdots j_{2p}]}%
K^{j_1}_{i_1}K^{j_2}_{i_2}\left( \frac{1}{2}\bar R^{j_3j_4}_{i_3i_4}-\epsilon s^2K^{j_3}_{i_3}%
K^{j_4}_{i_4} \right)\times \cdots\times \Bigg( \frac{1}{2}\bar R^{j_{2p-1}j_{2p}}_{i_{2p-1}i_{2p}} \notag \\
&& -\epsilon s^2K^{j_{2p-1}}_{i_{2p-1}}K^{j_{2p}}_{i_{2p}}\Bigg)-p\epsilon\int_0^1\!ds\,\delta^{[ii_2\cdots i_{2p}]}_{[j_1\cdots j_{2p}]}%
K^{j_1}_{j}K^{j_2}_{i_2}\left( \frac{1}{2}\bar R^{j_3j_4}_{i_3i_4}-\epsilon s^2K^{j_3}_{i_3}%
K^{j_4}_{j_4} \right)\times \cdots \notag \\
&& \cdots \times \Bigg( \frac{1}{2}\bar R^{j_{2p-1}j_{2p}}_{i_{2p-1}i_{2p}}-%
\epsilon s^2K^{j_{2p-1}}_{i_{2p-1}}K^{j_{2p}}_{i_{2p}}\Bigg)-\frac{1}{2}\delta^{[ii_1\cdots i_{2p-1}]}_{[jj_1\cdots j_{2p-1}]}\bar{\nabla}_{i_1}\bigg(K^{j_2}_{i_2}\bar{\nabla}^{j_1}K^{j_3}_{i_3}R^{j_4j_5}_{i_4i_5}\times\cdots \notag \\
&& \cdots\times R^{j_{2p-2}j_{2p-1}}_{i_{2p-2}i_{2p-1}}\bigg)-\frac{1}{2^{p+1}}\delta^{[ii_1\cdots i_{2p}]}_{[jj_1\cdots j_{2p}]}R^{j_1j_2}_{i_1i_2}\times\cdots \times R^{j_{2p-1}j_{2p}}_{i_{2p-1}i_{2p}}+\frac{\partial_w\pi^i_j}{N\sqrt{|h|}}\,,
\end{eqnarray}
or as a functional of intrinsic quantities as 
\begin{eqnarray} \label{EijL}
\mathcal{E}^i_{(p)j}&=&\frac{\partial_w\pi^i_j}{N\sqrt{|h|}}-p\epsilon\int_0^1\!ds(1-s)\delta^{[ii_1\cdots i_{2p}]}_{[jj_1\cdots j_{2p}]}%
K^{j_1}_{i_1}K^{j_2}_{i_2}\Big( \frac{1}{2}\bar R^{j_3j_4}_{i_3i_4}-\epsilon s^2K^{j_3}_{i_3}%
K^{j_4}_{i_4} \Big)\times \cdots \notag \\
&&\quad \cdots \times \Bigg( \frac{1}{2}\bar R^{j_{2p-1}j_{2p}}_{i_{2p-1}i_{2p}}-%
\epsilon s^2K^{j_{2p-1}}_{i_{2p-1}}K^{j_{2p}}_{i_{2p}}\Bigg)-p\epsilon\int_0^1ds\delta^{[ii_2\cdots i_{2p}]}_{[j_1\cdots j_{2p}]}%
K^{j_1}_{j}K^{j_2}_{i_2}\Big( \frac{1}{2}\bar R^{j_3j_4}_{i_3i_4}\notag \\
&&\qquad -\epsilon s^2K^{j_3}_{i_3}%
K^{j_4}_{i_4} \Big)\times \cdots\times \Bigg( \frac{1}{2}\bar R^{j_{2p-1}j_{2p}}_{i_{2p-1}i_{2p}}-%
\epsilon s^2K^{j_{2p-1}}_{i_{2p-1}}K^{j_{2p}}_{i_{2p}}\Bigg)+\bar{\mathcal{E}}^i_{(p)j} \, \notag \\
&&\qquad \quad-\frac{1}{2}\delta^{[ii_1\cdots i_{2p-1}]}_{[jj_1\cdots j_{2p-1}]}\bar{\nabla}_{i_1}\bigg(K^{j_2}_{i_2}\bar{\nabla}^{j_1}K^{j_3}_{i_3}\Big(\bar{R}^{j_4j_5}_{i_4i_5}-2K^{j_4}_{i_4}K^{j_5}_{i_5}\Big)\times\cdots \notag \\ 
&&\hspace{160pt} \cdots \times \Big(\bar{R}^{j_{2p-2}j_{2p-1}}_{i_{2p-2}i_{2p-1}}-2%
K^{j_{2p-2}}_{i_{2p-2}}K^{j_{2p-1}}_{i_{2p-1}}\Big)\bigg)\,.
\end{eqnarray}

\section{Additional identities}

We need to relate Kronecker deltas that differ in rank.
For a rank-four Kronecker delta, useful identities are
\begin{equation} \label{KKKK}
\delta^{[j_1\cdots j_4]}_{[i_1\cdots i_4]}K^{i_1}_{j_1}K^{i_2}_{j_2}K^{i_3}_{j_3}K^{i_4}_{j_4}= \delta^{[j_1j_2j_3]}_{[i_1i_2i_3]}\left (KK^{i_1}_{j_1}K^{i_2}_{j_2}K^{i_3}_{j_3}-3K^{i_1}_{j_1}K^{i_2}_{j_2}K^{i_3}_{l}K^{l}_{j_3}\right )\,,
\end{equation}
and
\begin{eqnarray} \label{KKR}
\delta^{[j_1\cdots j_4]}_{[i_1\cdots i_4]}K^{i_1}_{j_1}K^{i_2}_{j_2}\bar R^{i_3i_4}_{j_3j_4}&=& \delta^{[j_1j_2j_3]}_{[i_1i_2i_3]}\left (KK^{i_1}_{j_1}\bar R^{i_2i_3}_{j_2i_3}-K^{i_1}_{l}K^{l}_{j_1}\bar R^{i_2i_3}_{j_2j_3}-2K^{i_1}_{j_1}K^{i_2}_{l}\bar R^{li_3}_{j_2j_3}\right )\,.
\end{eqnarray}
Notice that the identity holds for any pair of tensors that share the same symmetries as the extrinsic and intrinsic curvature.
The generalization of the relations (\ref{KKKK}) and (\ref{KKR}) for $2m$ extrinsic curvatures and
$n-m$ Riemann tensors is

\begin{equation} \label{K2m}
 \begin{aligned}
  \delta^{[i_1\cdots i_{2n}]}_{[j_1\cdots j_{2n}]}K^{j_1}_{i_1}\cdots & K^{j_{2m}}_{i_{2m}}%
  \bar R^{j_{2m+1}j_{2m+2}}_{i_{2m+1}i_{2m+2}}\cdots \bar R^{j_{2n-1}j_{2n}}_{i_{2n-1}i_{2n}}=\\
  & \delta^{[i_1\cdots i_{2n-1}]}_{[j_1\cdots j_{2n-1}]}K^{j_1}_{i_1}\cdots K^{j_{2m-2}}_{i_{2m-2}}%
  \bar R^{j_{2m-1}j_{2m}}_{i_{2m-1}i_{2m}}\cdots \bar R^{j_{2n-5}j_{2n-4}}_{i_{2n-5}i_{2n-4}}\Big(%
  KK^{j_{2n-3}}_{i_{2n-3}}\bar R^{j_{2n-2}j_{2n-1}}_{i_{2n-2}i_{2n-1}} &\\
  & -(2m-1)K^{j_{2n-3}}_lK^l_{i_{2n-3}}\bar R^{j_{2n-2}j_{2n-1}}_{i_{2n-2}i_{2n-1}}%
  -(2m-2j)K^{j_{2n-3}}_{i_{2n-3}}K^{j_{2n-2}}_l\bar R^{lj_{2n-1}}_{i_{2n-2}i_{2n-1}}\Big)
 \end{aligned}
\end{equation}
where we factored out $2m-2$ extrinsic curvatures and $n-m-1$ Riemann tensors. 

In presence of a pair of free indices, we have
\begin{equation} \label{K2mij}
 \begin{aligned}
  \delta^{[ii_1\cdots i_{2n}]}_{[jj_1\cdots j_{2n}]}&K^{j_1}_{i_1}\cdots  K^{j_{2m}}_{i_{2m}}%
  \bar R^{j_{2m+1}j_{2m+2}}_{i_{2m+1}i_{2m+2}}\cdots \bar R^{j_{2n-1}j_{2n}}_{i_{2n-1}i_{2n}}=\\
  & \delta^{[ii_1\cdots i_{2n-1}]}_{[jj_1\cdots j_{2n-1}]}K^{j_1}_{i_1}\cdots K^{j_{2m-2}}_{i_{2m-2}}%
  \bar R^{j_{2m-1}j_{2m}}_{i_{2m-1}i_{2m}}\cdots \bar R^{j_{2n-5}j_{2n-4}}_{i_{2n-5}i_{2n-4}}\Big(%
  KK^{j_{2n-3}}_{i_{2n-3}}\bar R^{j_{2n-2}j_{2n-1}}_{i_{2n-2}i_{2n-1}} &\\
  & -(2m-1)K^{j_{2n-3}}_lK^l_{i_{2n-3}}\bar R^{j_{2n-2}j_{2n-1}}_{i_{2n-2}i_{2n-1}}%
  -(2m-2j)K^{j_{2n-3}}_{i_{2n-3}}K^{j_{2n-2}}_l\bar R^{lj_{2n-1}}_{i_{2n-2}i_{2n-1}}\Big)\\
 & \hspace{80pt} -\delta^{[ii_2\cdots i_{2n}]}_{[j_1j_2\cdots j_{2n}]}%
 K^{j_1}_{j}K^{j_2}_{i_2}\cdots K^{j_{2m}}_{i_{2m}} \bar R^{j_{2m+1}j_{2m+2}}_{i_{2m+1}i_{2m+2}}%
 \cdots \bar R^{j_{2n-1}j_{2n}}_{i_{2n-1}i_{2n}}\,,
 \end{aligned}
\end{equation}
that has one extra term --the last one-- in comparison to Eq. (\ref{K2m}). Notice that
we fixed $i_1$ when taking the trace to lower the degree of the generalized Kronecker symbol.

\end{document}